\documentclass[12pt]{iopart}
\usepackage{graphicx}

\begin{document}
\title{After-gate attack on a quantum cryptosystem}

\author{C. Wiechers$^{1,2,3,\dagger}$, L. Lydersen$^{4,5.\dagger}$, C. Wittmann$^{1,2,*}$, D. Elser$^{1,2}$, J. Skaar$^{4,5}$, Ch. Marquardt$^{1,2}$, V. Makarov$^{4}$ and G. Leuchs$^{1,2}$}

\address{$^1$ Max Planck Institute for the Science of Light, G\"unther-Scharowsky-Stra\ss e~1, Bau~24, 91058, Erlangen, Germany\\ $^2$ Universit\"at Erlangen-N\"urnberg, Staudtstra\ss e~7/B2, 91058, Erlangen, Germany \\ 
$^3$ Departamento de F\'isica, Campus Leon, Universidad de Guanajuato, Loma del Bosque 103, Fracc. Lomas del Campestre, 37150, Leon, Gto, Mexico \\
$^4$ Department of Electronics and Telecommunications, Norwegian University of Science and Technology, NO-7491 Trondheim, Norway \\
$^5$ University Graduate Center, NO-2027 Kjeller, Norway \\
$^\dagger$ These authors contributed equally to this work.}
\ead{$^*$Christoffer.Wittmann@mpl.mpg.de}

\begin{abstract}
We present a method to control the detection events in quantum key distribution systems that use gated single-photon detectors. We employ bright pulses as faked states, timed to arrive at the avalanche photodiodes outside the activation time. The attack can remain unnoticed, since the
faked states do not increase the error rate \emph{per se}. This allows for an intercept-resend attack, where an eavesdropper transfers her detection events to the legitimate receiver without causing any  errors. As a side effect, afterpulses, originating from accumulated charge carriers in the detectors, increase the error rate. We have experimentally tested detectors of the system id3110 (Clavis2) from ID~Quantique. We identify the parameter regime in which the attack is feasible despite the side effect. Furthermore, we outline how simple modifications in the implementation can make the device immune to this attack.
\end{abstract}

\maketitle

\section{Introduction\label{sec:intro}}

An intriguing feature of quantum optics is that it enables communication protocols which are impossible to achieve by classical means. One prominent example is quantum key distribution (QKD) \cite{Gis1,Sca1} which in principle allows two parties (Alice and Bob) to communicate with unconditional security. It is thus impossible for an arbitrarily powerful eavesdropper (Eve) to obtain knowledge of the transmitted information.

In the well-known Bennett-Brassard 1984 (BB84) protocol in its original form \cite{Ben1}, Alice sends single photons of different polarizations to Bob. Under ideal conditions, the security of this protocol can be rigorously proved \cite{May1}. Furthermore, practically feasible procedures for distilling a secret key from the exchanged quantum states are known \cite{Sho1}. During the distillation, Alice and Bob generate a key sequence out of their raw data stemming from the quantum state exchange.
Eve's attempt to gain knowledge results in a perturbation of the quantum states, such that her information about the raw key can be upper bounded. Alice and Bob can thus shrink their raw data such that Eve's knowledge of the resulting key sequence becomes negligible.

Rigorous security proofs show that Eve cannot successfully attack an ideal implementation of BB84. However, real implementations always exhibit deviations from the ideal model. In order to guarantee secure communication, such deviations must be included into the security proofs. One example is the use of weak coherent states instead of single photons which is considered in the Gottesmann-Lo-L\"utkenhaus-Preskill (GLLP) security proof \cite{Got1}. The resulting reduction of the key rate can be mitigated by modifications to the protocol such as in the decoy state method \cite{Hwa1,Wan1,Lo1} or in the Scarani-Acin-Ribordy-Gisin 2004 (SARG04) protocol \cite{Sca2}. More subtle deviations can result in side channels through which information can unnoticeably leak to Eve. For example, photons might carry information in unwanted degrees of freedom \cite{Nau1}. Once such side channels are known, they need to be considered in a more general security proof. 

Nowadays, quantum cryptography has matured to the point that several commercial products are available \cite{idQ,MaQ}. Each system might have loopholes which are particular to its implementation. Some implementations are, for example, susceptible to non-conforming light pulses that Eve sends into Alice's or Bob's devices. Eve could use reflectometry to read modulator states \cite{Vak1} or take control over the detectors by sending faked states \cite{Vad1,Vad2}, time-shifted pulses \cite{Zha1}, or by detector blinding combined with faked states \cite{Lar1}. The impact of such interventions strongly depends on the particular implementation. It is thus difficult to include them in general security proofs. Alternatively, specific countermeasures could be devised by adapting hard- or software of the systems, such that all assumptions in the security proof about the QKD module are again valid.

In this paper, we investigate a particular attack on the QKD device id3110 Clavis2 from  ID~Quantique. The fibre-based system utilizes the plug\,\&\,play principle \cite{Stu1} where the quantum states are encoded as relative phase of two pulses. In our experiment, we send irregular, bright light pulses (faked states) outside the activation time of the gated detectors. We show that we can generate measurement results in the Bob module with only a slight increase in the quantum bit error rate (QBER), if side effects of the attack are considered properly.

The paper is organized as follows. Section \ref{sec:2} describes the basic principles of our attack. In Section \ref{sec:detec}, we elaborate on the particular implementation of the detectors in the Clavis2 system. In Section \ref{sec:loophole}, we present the imperfections found in the system. Section \ref{sec:sideff} discusses the side effect of the faked-state attack, which actually partly protects the security of the system. Section \ref{sec:qber} presents all the necessary elements for simulations and shows the parameters for which the Clavis2 system is \textit{not} secure. In Section~\ref{sec:countermeasures}, we discuss possible countermeasures against the proposed attack before concluding in Section 8.

\section{Intercept-resend attack using faked states\label{sec:2}}

In the BB84 protocol~\cite{Ben1}, Alice randomly chooses one of two non-orthogonal bases to encode her quantum bit. Bob independently chooses his measurement basis at random. If his basis matches Alice's, he will measure the quantum state correctly. In half of the cases, however, Bob chooses the wrong basis. Alice and Bob compare encoding and measurement basis via a classical authenticated channel and remove all events with basis mismatch from their raw data.

In an intercept-resend attack, Eve places a copy of Bob's apparatus into the quantum channel. Then she performs the same kind of measurement as Bob, tries to reproduce the original quantum state and sends it to Bob. Since Eve is unaware of Alice's basis choice, she will inevitably introduce errors in case of a basis mismatch between her basis and the one used by Alice and Bob. Eve will thus always be detected in a perfect implementation of a QKD system~\cite{Sho1,Got1}.

In case of an imperfect implementation, however, Eve may attack the QKD system by sending faked states instead of quantum states \cite{Vad1}. Her aim is to generate faked states which only produce a detection event in the Bob module if Eve's basis matches Bob's basis. In this case, after Alice and Bob discard their non-matching bases, all that remains in the key are bits for which Alice, Eve and Bob had the same basis. Thus, Eve generates no errors.

After the attack, Bob and Eve share identical bit values and basis choices. The attack works on widely used QKD protocols, namely BB84, SARG04, and the decoy method. The attack exhibits an extra 3 dB loss because of the possible basis mismatch of Eve and Bob. This is easily compensated in a practical Eve, since she may use better detector efficiencies and exclude loss in the line~\cite{Vad1}.

\section{Detectors in Clavis2\label{sec:detec}}

\begin{figure}
\centering
\includegraphics[width=12cm]{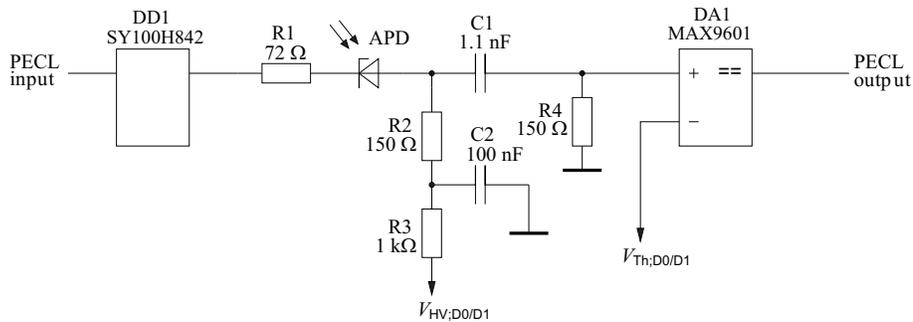}
\caption{\label{fig:circ} Equivalent circuit diagram of Bob's detectors in Clavis2. See text for description.}
\end{figure}

The impact of faked states strongly depends on the implementation of the detection scheme in a QKD system. We will focus on systems employing avalanche photodiodes (APDs) in the Geiger mode, as it is the case in many QKD systems \cite{Rib1,Tri1}, and all commercially available realizations \cite{idQ,MaQ}. Furthermore, we assume that the APDs are gated, {i.e.} activated only in time intervals when signal states are expected to arrive. During the activation time, a large reverse voltage is applied to the APDs such that the APDs are biased above the breakdown voltage. Then a single photon can trigger a carrier avalanche which results in a macroscopic current. If the generated current exceeds a certain threshold, a detection event (click) is registered. 

As an example, we consider the behavior of the gated detectors in ID~Quantique's Clavis2 QKD system. A detector circuitry reverse-engineered by us is shown in Figure~\ref{fig:circ}. The APDs are biased by the high voltage supply with $V_\mathrm{HV;D0/D1}$ almost as large as the breakdown voltage ($V_{\rm{HV};D0} = {-} 42.89\,\mathrm{V}$ and $V_{\rm{HV};D1} = {-}43.08\,\mathrm{V}$). The detectors are gated in Geiger mode by means of TTL signals, which are applied on top of the bias voltage with a period of 200 ns. The gates are supplied as PECL logic level signals from the main board and converted to TTL signals by the buffer DD1. The comparator DA1 monitors the APD current and registers a click in the detector when the current peak passes a threshold ($V_{\rm{Th};D0} = 77\,\mathrm{mV}$, $V_{\rm{Th};D1} = 84\,\mathrm{mV}$). The comparator produces a PECL output pulse for each detection event.

During all the time not covered by the gate, each APD is biased at a constant value $V_{\rm{HV};D0/D1}$ below the breakdown voltage. The current through the APD is then approximately proportional to the incident optical power. The circuit behaves similarly as a linear photodiode followed by a comparator.

\section{Description of loopholes in the system \label{sec:loophole}}

\begin{figure}
\centering
\includegraphics[width=8cm]{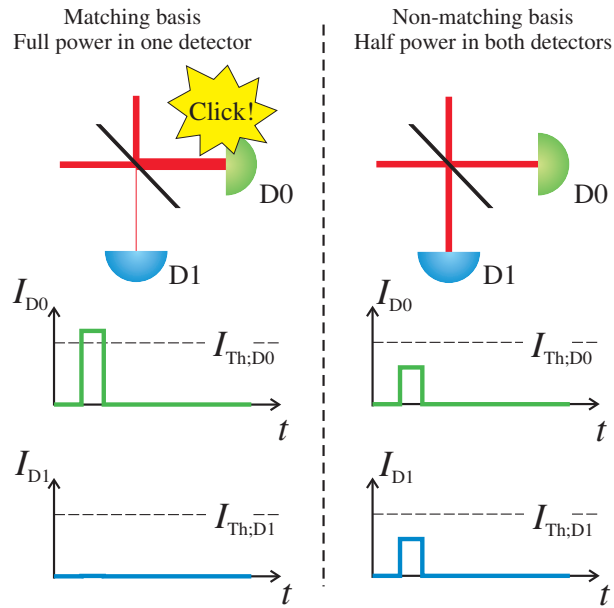}
\caption{\label{fig:attack}
Principle of detector control. In the detection part of a phase-encoded QKD, the two pulses which could be generated by Eve as a faked state interfere at a 50:50 beamsplitter. (a) For Bob's basis choice matching Eve's, the signals interfere such that detector D0 clicks deterministically, because the photocurrent surpasses the detection threshold $I_\mathrm{Th;D0}$. (b) For Bob's basis choice not matching Eve's, the power is split 50:50 to both detectors. The photocurrent does not surpass the threshold. Therefore, the faked state is not detected.}
\end{figure}

In the following subsections, we describe two unexpected deviations of the detection system from the idealized behavior implicitly assumed by the designers of the QKD system. We start by explaining the detection process in detail.
In an ideal plug\,\&\,play system, the relative phase between the signal states and reference pulses in the receiver module ($0,\pi /2,\pi,3\pi /2$) is determined by a combined phase modulation of Alice and Bob, {i.e.} by a combination of Alice's bit and basis ($0,\pi /2,\pi,3\pi /2$) and Bob's measurement basis ($0,\pi /2$).

Let us consider a standard intercept-resend attack. For a matching basis choice of Alice, Eve and Bob, the phase difference is $0$ or $\pi$. This restricts the possible outcome of the measurement to a single detector and results in a conclusive outcome for  Bob. For a mismatched basis choice, the phase difference is $\pi /2$ or $3\pi /2$. In this case, either of their detectors will click randomly. This clearly causes a QBER of 25\%. 

\subsection{Linear mode APDs}

In the linear regime of the APDs, Eve can substitute the quantum states with bright coherent states~\cite{Lar1}.  Figure~\ref{fig:attack} shows examples of pulses which generate a click only if Bob's and Eve's bases match, since the comparator following the APD will only click, if the input optical power surpasses a critical power threshold. In case of a basis mismatch, the optical power is distributed equally among the detectors and no detection click is generated. 

To exploit the loophole experimentally, we look closer at the detector characteristics. As mentioned, the APDs are biased below the breakdown voltage before and after the gate. Optically, Bob's phase modulation extends temporally on either side of the gate pulse by approximately 10\,ns. We have verified that the system accepts clicks at least 10.5 ns after the gate, still assigning the click to the bit slot associated with the gate. 

\begin{figure}
\centering
\begin{tabular}{l}
\includegraphics[width=14cm]{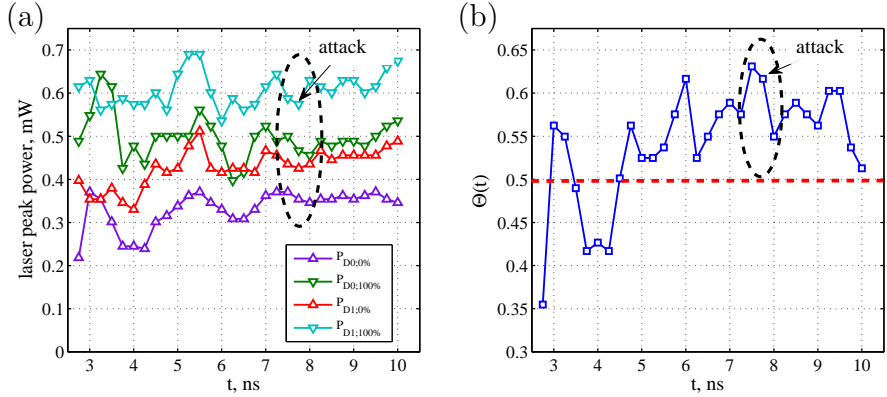}\\[-5.6cm]
\hspace{1cm}(a) \hspace{5.2cm} (b)\\[5.0cm]
\end{tabular}
\caption{
\label{fig:thres} Detection click thresholds in Clavis2 for a pulse duration of 0.12\,ns. (a) Power thresholds $P_\mathrm{D0,D1;0\%,100\%}$ for 0\,\% and 100\,\% probability of a bright pulse detection in detector D0 and D1. The fluctuations are reproducible and are probably caused by the fluctuating bias voltage after the gate. (b) Calculated $\Theta (t)$ (see Eq.~\ref{equ:pa}), which shows that an atttack is possible for delays of 4.5\,ns to 10\,ns  with an optimal and comfortable margin of $\Theta$ at $7.5\,$ns.}
\end{figure}

We sent bright laser pulses to both detectors before and after they are gated, in order to find the click thresholds of each detector. A perfect control of Bob is possible, if the maximum power at which the detectors do not produce clicks is not lower than half the power at which they always produce a click. This can be written as
\begin{eqnarray}
\Theta(t)=\frac{\texttt{min}\left\{ P_{\rm{D0};0\%}\left( t \right), P_{\rm{D1};0\%}\left( t \right) \right\}}{\texttt{max}\left\{ P_{\rm{D0};100\%}\left( t \right), P_{\rm{D1};100\%}\left( t \right) \right\} }> 0.5, \label{equ:pa}
\end{eqnarray}
where $t$ is the time between the leading edge of the gate and the bright pulse, $P_{\rm{D0};0\%}(t)$ is the maximum power that does not generate a click in D0 and $P_{\rm{D0};100\%}(t)$ is the minimum power that certainly generates a click in D0 (analogously for D1).

We have found that the linear behavior prior to the gate can not be exploited, since charge carrier generation results in a large afterpulse effect during the gate.  For an attack after the gate, Figure \ref{fig:thres} shows the experimentally measured power thresholds and the corresponding values of $\Theta(t)$ for 0.12\,ns long 1550\,nm laser pulses. The figure shows that an attack is feasible in a wide time window with the maximum value of $\Theta(t)$ at 7.5 ns after the gate. At this time, a 587\,$\mathrm{\mu W}$ laser pulse can cause a click in both detectors, while a 293.5\,$\mu$W laser pulse will never cause a click in any detector. This result reveals a weak spot in the system. We have found, however, that the attack can not be applied straightforwardly, because of an afterpulsing side effect, which is discussed in Section~\ref{sec:sideff}. Therefore we attacked at the point 7.75 ns after the gate to slightly reduce the maximum laser power applied to the system. At 7.75 ns after the gate, a 575\,$\mathrm{\mu W}$ laser pulse can cause a click in both detectors, while a 287.5\,$\mu$W laser pulse will never cause a click in any detector.

\subsection{Faked states applied during the dead time\label{no_dead}}

As a second loophole in the system, we have found that the system registers detection events from bright faked states at any time.  Typically, the device applies a dead time of 10\,$\mu$s whenever the system registers a click at any of the detectors, not gating both APDs for the duration of the dead time~\cite{Vad3}. However, we have found that the time between the detection events originating from our faked states can be as short as 30\,ns.

\begin{figure}
\centering
\includegraphics[width=12cm]{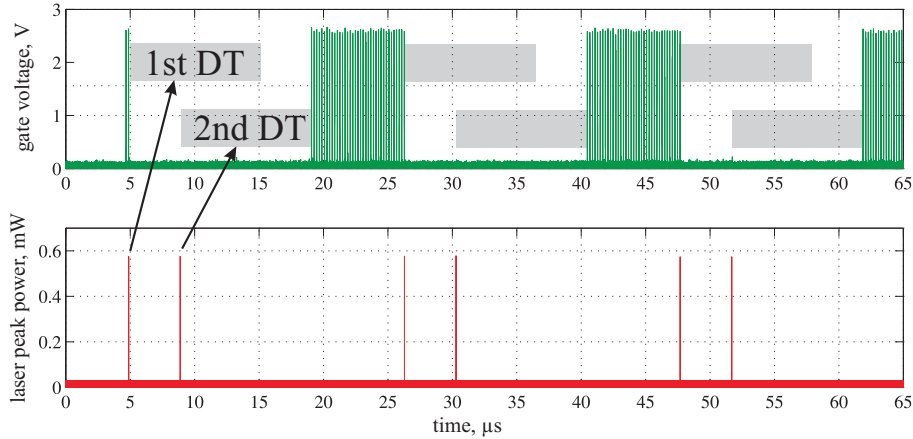}
\caption{\label{fig:pls2} Dead time extension behavior. The figure shows that if a bright pulse (or avalanche) causes a dead time (1st DT), any bright pulse during the dead time will be a valid detection and causes an extra dead time (2nd DT). Therefore it extends the effective dead time. (upper oscillogram) The gate pattern applied to the detector. (lower oscillogram) Optical power of successive bright pulses impinging on the detector with a delay of 4\,$\mu$s.}
\end{figure}

Figure~\ref{fig:pls2} shows the effect of a bright pulse arriving during the dead time. The electronic logic registers a valid click and subsequently resets the dead time to another 10\,$\mu$s after the second bright pulse. We found experimentally that in the deadtime all faked states with laser peak power of 575\,$\mu$W were detected by detector D0 while the detection probability of Bob's D1 was $\eta_\mathrm{B} > 0.99985$. In Section \ref{sec:strat}, we will show how this loophole can be exploited in order to overcome the negative side effect of afterpulses, which is described in the next section.

\section{Characterization of afterpulsing side effect \label{sec:sideff}}

Once a detection is registered in a gated APD, a long dead time is typically applied to reduce afterpulsing. This dead time is considerably longer than the inverse of the gating frequency and is typically of the order of several microseconds.

The afterpulse effect is due to carrier traps, which are populated by avalanche currents in the detection process~\cite{Tri1, Cov1}.
We have found that bright pulses also populate the carrier traps, irrespective of whether they generate detection events or not. Without a registered detection, a dead time is not applied  by the detectors circuitry. The carriers released from traps can therefore cause afterpulsing in the detector. These uncontrollable clicks will contribute to the QBER.

\begin{figure}
\centering
\includegraphics[width=9 cm]{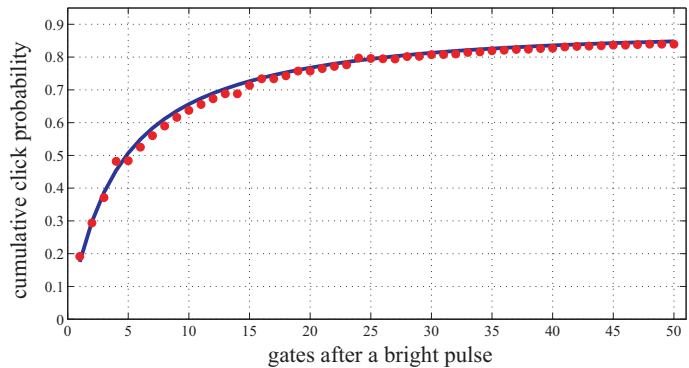}
\caption{\label{fig:acumfig} Afterpulses caused by the after-gate attack. The chart shows the experimentally measured cumulative probability to get at least one dark count after a 287.5\,$\mu$W pulse applied to both detectors (red dots), a Monte Carlo simulation of the same process using the parameters from Table~\ref{tab:decpar} (solid line). }
\end{figure}

We have characterized this side effect of the after-gate attack in the successive gates by plugging a laser directly to one of the fiber inputs of the 50:50 beamsplitter of Fig.~\ref{fig:attack}. The laser pulses have a peak power of 287.5\,$\mu$W for each detector. As expected, the pulse never causes a click immediately. However, very often it causes an afterpulse within the following gates. Figure \ref{fig:acumfig} shows the cumulative probability to get a click in any of the two detectors in the next gates. After 50 gates, the cumulative probability to get a random click has reached 84 \%, which could jeopardize Eve's attack by causing a to high QBER.

Note that the system sends frames of 1075 pulses as dictated by the send-return configuration~\cite{Stu1}. Therefore, the attack can always be applied in the end of the frame with a reduced risk of a random afterpulse. If the system requires only one detection every second frame, then the security is completely compromised. Additionally, the attack may be applicable for a different set of system parameters, e.g. different operation frequencies of Bob.

We have modeled the afterpulse effects of carrier traps. We have found that the probabilities $P_{\rm{ap};D0/D1}\left(t_j\right)$ of a detection event after a faked-state attack can be modelled using a double exponential decay for the detectors:
\begin{equation}
P_{\rm{ap};D0/D1}\left(t_j\right)=P_{\rm{dark};D0/D1}+(1-P_{\rm{dark;D0/D1}})\sum^{2}_{i=1}A_{i\mathrm{;D0/D1}}\textbf{e}^{-t_j/\tau_{i\mathrm{;D0/D1}}}, \label{equ:dobdec}
\end{equation}
where $P_{\mathrm{dark};D0/D1}$ is the dark count probability, $A_{i\mathrm{;D0/D1}}$ are probability amplitudes that depend on the number of carriers which are generated in the detector, and $\tau_{i\mathrm{;D0/D1}}$ are the associated decay constants. The afterpulse probabilities in Fig.~\ref{fig:acumfig} were reproduced by a Monte Carlo simulation using the double exponential decay model given by Eq.~\ref{equ:dobdec}. By iterating the Monte Carlo simulation, the decay parameters were found by minimizing the squared distance between the measurement data and the simulation data, equivalent to the method of least squares in regression analysis. Table~\ref{tab:decpar} shows the resulting decay parameters, and the final Monte Carlo simulation is shown in Fig.~\ref{fig:acumfig}. The decay parameters are in agreement with earlier published data on APDs~\cite{Tri1,Cov1}.

\begin{table}
\begin{indented}
\item[]\begin{tabular}{@{}*{4}{c}}
\br
 \centre{2}{ \emph{Detector 0}}   &  \centre{2}{ \emph{Detector 1}}     \\
   \emph{Parameter} & \emph{Value} & \emph{Parameter} & \emph{Value} \\ 
\mr
   $P_{\rm{dark;D0}}$ & $1.158\cdot10^{-4}$  &    $P_{\rm{dark;D1}}$ & $3.812\cdot10^{-4}$ \\ 
   $A_{1;\rm{D0}}$ & $3.572\cdot10^{-2}$    & $A_{1;\rm{D1}}$ & $1.068\cdot10^{-1}$ \\ 
   $A_{2;\rm{D0}}$ & $2.283\cdot10^{-2}$   & $A_{2;\rm{D1}}$ & $5.054\cdot10^{-2}$   \\ 
   $\tau_{1;\rm{D0}}$ & $1.159\,\mathrm{\mu s}$    & $\tau_{1;\rm{D1}}$ & $0.705\,\mathrm{\mu s}$   \\ 
   $\tau_{2;\rm{D0}}$ & $4.277\,\mathrm{\mu s}$  &    $\tau_{2;\rm{D1}}$ & $3.866\,\mathrm{\mu s}$ \\ 
\br
\end{tabular}
\end{indented}
\caption{\label{tab:decpar}Decay parameters of trap levels in both detectors. These parameters were used for the Monte Carlo simulation shown in Fig.~\ref{fig:acumfig}. }

\end{table}

\section{Simulations of after-gate attack and QBER estimation \label{sec:qber}}

We estimate the QBER for different attack scenarios using a Monte Carlo simulation. In our simulation, Alice and Bob use the BB84 protocol. Eve performs a faked-state attack by putting her modified Bob and Alice modules in the channel. Eve places her Bob module in the beginning of the line next to Alice.  We assume that Alice sends an optimized signal amplitude~\cite{Bra1} where the sent mean photon number $\mu$ is equal to the channel transmittance $T$. Unless otherwise noted, Eve measures this signal with perfect detectors (100\,\% efficiency and noiseless) and a lossless aparatus. Then she reproduces a bright faked state with the corresponding bit value for Bob.

Bob's module is simulated including realistic parameters which were determined experimentally for our device. Besides the parameters for the afterpulsing and dark count effects (see Table~\ref{tab:decpar}), there are the optical transmittance of Bob's setup ($T_B=0.412$), the quantum efficiency of the detectors ($\eta_B=0.1$) and the detector dead time ($\tau_\mathrm{dead}=10\,\mu$s).

In the simulation, we process the consecutive gates of a frame separately. We incorporate the side effect by increasing the afterpulse probability of a detector, if carriers were generated either by a regular avalanche or by bright pulses with full or half power\footnote{Carrier generation by the half-power pulses is the most important effect, because the system does not apply the deadtime after them.}. We have experimentally verified that for the operation frequency and used optical powers the carrier traps in the detectors are not saturated by our attack and that the afterpulses of the two carrier-generating processes  with different lifetimes occur independently and with Poissonian statistics~\cite{Goodman}. The afterpulse probability of a gate at time $t_j$ is then increased by a previous gate with carrier generation at time $t_k$ as
\begin{eqnarray}
P^\mathrm{new}_{\rm{ap};D0/D1}(t_j)&=&P^\mathrm{old}_{\rm{ap};D0/D1}(t_j)+(1-P^\mathrm{old}_{\rm{ap};D0/D1}(t_j))\times\nonumber\\
&&\sum^{2}_{i=1}\gamma_{i\mathrm{;D0/D1}}A_{i\mathrm{;D0/D1}}\textbf{e}^{-(t_j-t_k)/\tau_{i\mathrm{;D0/D1}}}, \label{equ:accumsim}
\label{eq:incr_afterpulse}
\end{eqnarray}
where $\gamma_{i\mathrm{;D0/D1}}$ is a correction of the probability amplitude $A_{i\mathrm{;D0/D1}}$. In case of a bright pulse attack with 287.5\,$\mu$W pulses $\gamma_{i\mathrm{;D0/D1}}=1$. For a bright pulse attack with full 575\,$\mu$W power to one detector (successful attack), we increase the afterpulse probability twice applying Equation~\ref{eq:incr_afterpulse}. We have measured that a regular avalanche in D0 and D1 has $\{\gamma_{i\mathrm{;D0}},\gamma_{i\mathrm{;D1}}\}=\{1.836,3.673\}$.

\subsection{Strategy of Eve with dead time}

\begin{figure}
\centering
\includegraphics[width=8cm]{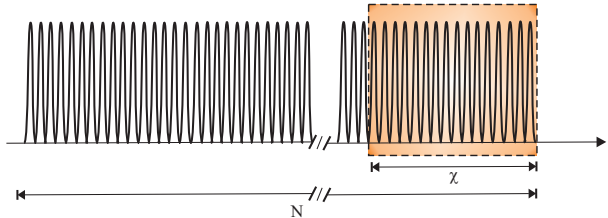}
\caption{\label{fig:xframe} We attack only the last $\chi$ gates of the total number of gates $N=1075$ in the frame, such that the raw key rate generated by the attack on each frame is equal to the rate without eavesdropping.}
\end{figure}

We first simulated the QBER without the deadtime loophole described in Section~\ref{no_dead}, {i.e.} assuming that Bob rejects detection events during the detector dead time. To increase the performance of Eve's attack, she adopts the following strategy: (i) Attack only the last $\chi$ gates of the total of $N$ gates of a frame, as shown in Figure~\ref{fig:xframe}. This will lead to a larger trapped carrier density at the end of the frame, which, when gates are absent, is ignored by the detectors. 
(ii) Use a small classical memory (up to three consecutive gates), which allows for checking whether she received several consecutive clicks. These are then sent to Bob as a \textit{burst attack}. This will lead to a decreased time between failed attacks and following attacks, which suppresses the afterpulsing by forcing earlier dead time. (iii) After the burst attack, wait as long as the dead time of Bob's detectors, in order to avoid carrier generation and afterpulsing directly after the dead time.

\begin{figure}
\centering
\begin{tabular}{l}
\hspace{2mm}\includegraphics[width=15cm]{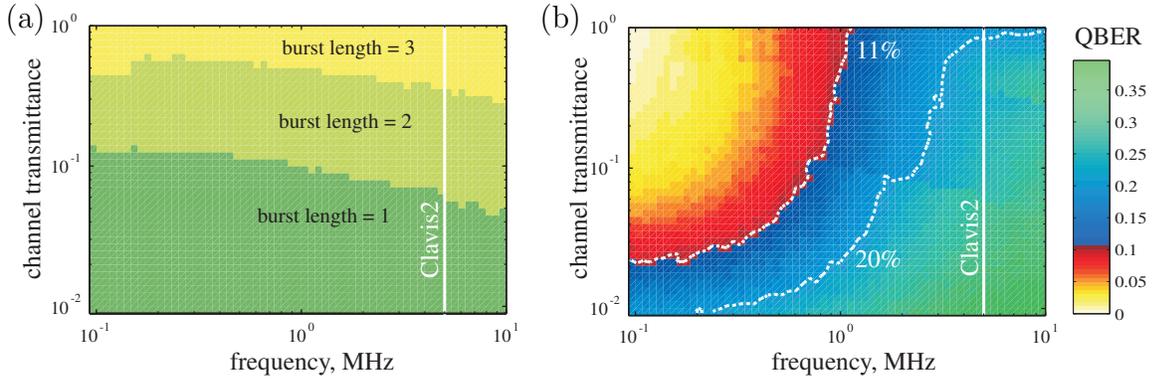}\\[-5.2cm]
(a) \hspace{6.3cm} (b)\\[4.5cm]
\end{tabular}
\caption{\label{fig:keep} Simulated attack performance for the case when Bob discards clicks during the detector dead time. (a) Burst length for different channel transmittances and gate frequencies. (b) QBER generated by the attack. We show contour lines for QKD security proofs which are more~\cite{Chau} or less~\cite{Sho1} tolerant to errors, allowing for a QBER of 20\,\% or 11\,\%, respectively.}
\end{figure}

We perform a simulation of the QBER induced by the attack for varying repetition rate and channel transmittance. The simulation consists of two major steps. Firstly, Eve adjusts the number of attacked gates $\chi$ in order to adjust the channel transmittance $T$ to the one anticipated by Alice and Bob. Eve tries to maximize the burst length in her attack. For a decreasing channel transmittance, Eve, however, receives fewer photons from Alice.  Therefore, the maximal burst length decreases for decreasing transmittance, see Fig.~\ref{fig:keep}(a).  Secondly, the QBER is simulated for $10^4$ frames. The average QBER is shown in Figure~\ref{fig:keep}(b) and compared to upper bounds of two different security proofs~\cite{Sho1, Chau}. The protocol in~\cite{Chau} would require a single photon source and is therefore not directly applicable in Clavis2. Therefore, we find that the attack can not compromise the security of Clavis2 due to increased afterpulse probability at the gate repetition rate of 5\,MHz. However, the security would be compromised for a more advanced system using single photons and the protocol in Ref.~\cite{Chau}, or for gate frequencies below about 1\,MHz. We note that there are numerous experimental setups and a commercial QKD system~\cite{MaQ} working below the critical operation frequency of about 1\,MHz. Additionally, technological improvements in the detectors could reduce the afterpulse effects and thereby enable the attack for high frequencies.

\subsection{Strategy of Eve without dead time\label{sec:strat}}

\begin{figure}
\centering
\begin{tabular}{l}
\hspace{2mm}\includegraphics[width=15cm]{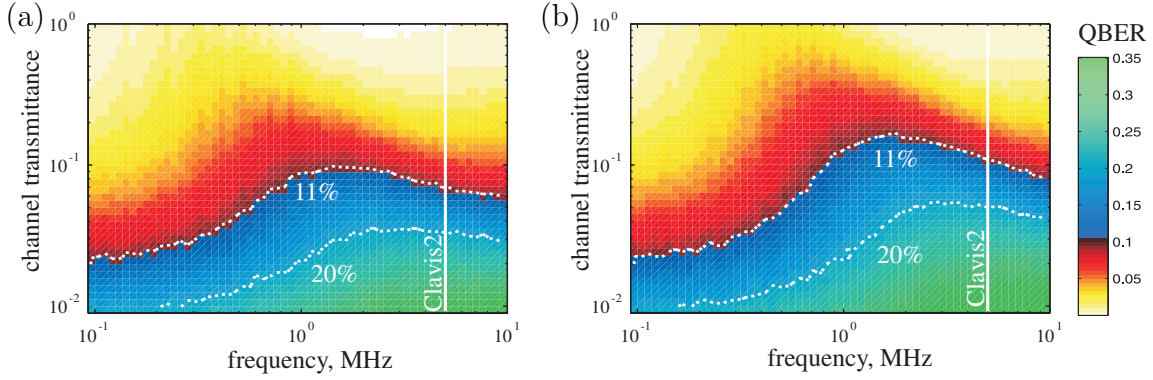}\\[-5.2cm]
(a) \hspace{6.3cm} (b)\\[4.5cm]
\end{tabular}
\caption{\label{fig:best} Simulated attack performance without dead time. The figures show the QBER generated by the attack taking advantage of the sensitivity of the detectors during the dead time together with the elongation of the dead time. (a) The QBER with a perfect Eve. The attack is feasible for all repetition rates and a wide range of channel transmittances. (b) QBER with a realistic Eve with a detection efficiency $T_B=0.5$ and a dark count probability $P_\mathrm{dark}=10^{-5}$, corresponding to a technically advanced but feasible eavesdropper.}
\end{figure}

Eve can adapt her attack strategy, if she has access to both the after-gate and dead-time loopholes. In the following, we show a strategy which is not an optimized one, but a rather intuitive and (as it turns out) successful approach. 
Eve again attacks the end of the frame as shown in Figure~\ref{fig:xframe}. Her strategy is to attack as frequently as possible. Thereby, she quickly enters a dead time of Bob's detectors. She will generate detection events during the dead time and, thereby, can prolong the detector dead time as shown in Section~\ref{no_dead}. Ideally, a major part of the attack happens during the dead state, which would completely remove the effect of afterpulses and result in negligible QBER for this part of the attack.

In the simulation, we again adjust the number of attacked gates $\chi$ and simulate the QBER for $10^4$ frames. Figure~\ref{fig:best} shows that for high transmittance the QKD system is vulnerable against the advanced attack, including for an eavesdropper with detection efficiency implementable today. The photon statistics are maintained during the attack. It is therefore applicable also to decoy state protocols.
  
\section{Countermeasures\label{sec:countermeasures}}
Note that both eavesdropping strategies (especially the latter one) leave strong fingerprints. In the latter case, the distance between two valid detection events can be smaller than the dead time of 10\,$\mu$s. Therefore, one countermeasure is to search for too closely timed detection events. Furthermore, rejecting detections during the dead time would restrict eavesdropping to lower frequencies as shown by our first simulation (see Figure~\ref{fig:keep}). A complete protection against the presented attacks is guaranteed if the detection times are resolved, such that Bob can discriminate between detections inside and outside the single-photon-sensitive part of the gate. Note however that this is highly non-trivial since the intrinsic jitter caused by the avalanche build-up is about equal to the length of the gate itself. Alternatively, a watchdog detector can be placed at Bob's input in order to detect bright faked states. Since such a detector cannot be an avalanche detector (this can be hacked), the countermeasure is only effective against bright faked states.

\section{Conclusions}

We have demonstrated that gated detectors in QKD systems can be controlled by an external eavesdropper using bright laser pulses during the linear mode operation. In particular, we have analysed the attack parameters for the commercial QKD system Clavis2 from ID~Quantique. In principle, the system is controllable by bright control pulses arriving after the gate time. However, we have found a side effect: afterpulse generation due to the faked states. The side effect generates high QBER, and therefore actually protects the system from a straightforward faked-state attack. Eve can, however, take advantage of a second imperfection, namely that the system accepts the bright pulses even in the dead time and, furthermore, resets the remaining dead time. In a simulation of the attack, we have found that the system is not secure if clicks during the dead time are accepted. The presented after-gate attack can be used independently or together with the blinding attack in Ref.~\cite{Lar1}. Although the after-gate attack in contrast to the blinding increases the QBER, it has the advantages that the optical power sent into the Bob module is weaker. Therefore, the after-gate attack is harder to detect with a watchdog detector. Another advantage is that this attack can be applied to detectors that are not blindable.

ID~Quantique has been notified about this loophole prior to the submission of the manuscript, and has implemented countermeasures.

\ack
We acknowledge ID~Quantique's valuable assistance in this project. We also thank Georgy Onishchukov and Nitin Jain for fruitful discussions. This work was supported by the Research Council of Norway (grant no{.} 180439/V30), DAADppp mobility program financed by NFR (project no{.} 199854) and DAAD (project no{.} 50727598). Ca{.}~Wi{.} acknowledges support from FONCICYT project no{.} 94142.

\end{document}